\documentclass[MSNbibl,number,seceqn,dvips]{arxstspdf}
\usepackage{url,breakurl}
\usepackage{graphicx}
\usepackage{flushend}
\usepackage{stfloats}

\volume{29}
\issue{4}
\pubyear{2014}
\firstpage{679}
\lastpage{686}
\doi{10.1214/14-STS478} 

\makeatletter

\newproclaim{example}{Example}[section]

\newcommand{\A}{\mathcal{A}}
\renewcommand{\b}{{\beta}}
\newcommand{\g}{{\gamma}}
\renewcommand{\H}{\mathcal{H}}
\newcommand{\m}{{\mu}}
\newcommand{\n}{{\nu}}
\newcommand{\PP}{\mathbb{P}}
\newcommand{\RR}{\mathbb{R}}
\renewcommand{\t}{{\tau}}
\newcommand{\UU}{\mathbb{U}}
\newcommand{\X}{\mathcal{X}}
\newcommand{\half}{{\frac{1}2}}
\newcommand{\thalf}{{\half}}
\newcommand{\E}{\mathrm{E}}
\renewcommand{\Pr}{\mathrm{P}}
\newcommand{\var}{\operatorname{var}}
\newcommand{\cov}{\operatorname{cov}}
\newcommand{\given}{\vert}
\newcommand{\Seen}{1}
\newcommand{\Sj}[1]{#1}
\makeatother

\begin{document}
\begin{frontmatter}

\title{Higher Order Tangent Spaces and Influence Functions}%
\runtitle{Higher Order Tangent Spaces and Influence Functions}

\begin{aug}
\author[a]{\fnms{Aad}~\snm{van der Vaart}\corref{}\ead[label=e1]{avdvaart@math.leidenuniv.nl}}
\runauthor{A. van der Vaart}

\affiliation{Leiden University}

\address[a]{Aad van der Vaart is Professor,
Mathematical Institute, Leiden University,
P. O. Box 9512, 2300 RA Leiden, The Netherlands \printead{e1}.}
\end{aug}

%
\begin{abstract}
We review higher order tangent spaces and influence functions
and their use to construct minimax
efficient estimators for parameters in high-dimensional semiparametric
models.
\end{abstract}

%
\end{frontmatter}

\section{Main Discussion}
\label{SectionIntroduction}

The concept of \emph{influence function} of an estimator was
originally coined in the theory of
robust statistics, and as an \emph{asymptotic influence function} played
a role in the development of semiparametric statistics
\mbox{(\cite{BKRW,vdVStFlour})}.
If an estimator $T_n$ of a quantity $\m$ based on a random sample of
observations $X_1,X_2,\ldots,X_n$ possesses an asymptotic expansion of
the form
%
%
\begin{equation}
\label{EqASLINEAR} T_n=\m+\frac{1} n{\sum
_{i=1}^n}\psi(X_i)+o_P
\bigl(n^{-1/2}\bigr),
\end{equation}
then the function $\psi$ is its \emph{asymptotic influence
function}. The name derives from the fact that if an observation
$X_i$ is replaced by a value $x$, then the change in the estimator
is $n^{-1} (\psi(x)-\psi(X_i) )$, at least if the
remainder term $o_P(n^{-1/2})$ is neglected.
The estimator is ``asymptotically robust'' if this change is
bounded in $x$, that is, if the influence function $\psi$ is bounded.

Semiparametric theory as developed in the 1980s/\break 90s was not concerned
with robustness, but with efficient estimation.
Provided that the variables $\psi(X_i)$ have zero mean and finite
variance, the expansion (\ref{EqASLINEAR}) implies that the sequence
$\sqrt n(T_n-\m)$ is asymptotically normally distributed with mean
zero. Among different asymptotically unbiased estimators,
the ones with small asymptotic variance are preferred. Semiparametric
lower bound theory showed that under so-called ``asymptotic regularity''
estimators with an expansion (\ref{EqASLINEAR}) with
$\psi$ the \emph{efficient influence function} attain the
smallest variance. Furthermore, it showed how to compute
the latter function from the \emph{tangent space} of the underlying
semiparametric model
(\cite{KoshevnikLevit,Pfanzagl82,BegunHallHuangWellner},
and \cite{vdV88}).

Higher order tangent spaces and influence functions are
generalizations of these concepts, but were developed by Robins et al.
\cite{RobinsetalFreedman}
from the perspective of constructing estimators
rather than asymptotic efficiency.
Thus, it will be fruitful to also give the definitions of influence
functions and tangent spaces from the point of view of constructing
estimators.

Assume that the observations $X_1,\ldots, X_n$ are
a random sample from a distribution $P_{\eta}$
with density $p_{\eta}$ relative to a measure $\m$
on a sample space $(\X,\A)$. The parameter
${\eta}$ is known to belong to a subset $\H$ of a normed
space, and it is desired to
estimate the value $\chi({\eta})$ of a functional $\chi\dvtx \H\to
\RR$.
Interest is in the situation of a semiparametric
or nonparametric model, where $\H$ is infinite-dimensional
and the dependence ${\eta}\mapsto p_{\eta}$ is assumed smooth (as in
\cite{vanderVaart91}).

Given a ``consistent'' initial estimator $\hat{\eta}$ of
${\eta}$, the ``plug-in estimator'' $\chi(\hat{\eta})$ is typically
consistent
for the parameter of interest $\chi({\eta})$, but it
may not be a good estimator. In particular, if $\hat{\eta}$
is a general purpose estimator, not specially constructed
to yield a good plug-in, then $\chi(\hat{\eta})$ will often
have a suboptimal precision.
To gain insight in this situation assume that the
parameter permits a Taylor expansion of the form
%
%
\begin{equation}
\label{EqTaylorChiLinear} \quad\chi({\eta})=\chi(\hat{\eta})+\chi_{\hat{\eta}}'({
\eta}-\hat {\eta}) +O \bigl(\|{\eta}-\hat{\eta}\|^2 \bigr).
\end{equation}
Such an expansion suggests that the plug-in estimator will have
an error of the order $O_P (\|{\eta}-\hat{\eta}\| )$, unless
the linear term $\chi_{\hat{\eta}}'({\eta}-\hat{\eta})$ in the
expansion vanishes and the error
has the square of this order. For a large parameter
set, the latter estimation error will typically be large.

The expansion (\ref{EqTaylorChiLinear}) also suggests that better estimators
can be obtained by ``estimating'' the linear term.
To achieve this assume a ``generalized
von Mises representation'' of the derivative of the form
%
%
\begin{eqnarray}
\label{EqVonMisesExpansionLinear}
\chi_{\hat{\eta}}'({\eta}-\hat{\eta}) &=&\int\dot
\chi_{\hat{\eta}}^\Seen\,d(P_{\eta}-P_{\hat{\eta}})
\nonumber
\\[-8pt]
\\[-8pt]
&=&P_{\eta}\dot\chi_{\hat{\eta}}^\Seen+O \bigl(\|{\eta}-\hat{
\eta }\|^2 \bigr),\nonumber
\end{eqnarray}
for some measurable function $\dot\chi_{\hat{\eta}}^\Seen\dvtx \X
\to
\RR$. Here
$Pf$ is short for the integral $\int f \,dP$, and it is assumed
that $P_{\eta}\dot\chi_{\eta}^\Seen=0$ for every ${\eta}$ [which
can always be
arranged by a recentering, as $\int1 \,d(P_{\eta}-P_{\hat{\eta}})=0$].
The von Mises representation (\ref{EqVonMisesExpansionLinear}) and
(\ref{EqTaylorChiLinear})
suggest the ``corrected plug-in estimator''
%
%
\begin{equation}
\label{EqPlugInEstimatorLinear} T_n=\chi(\hat{\eta})+\PP_n \dot
\chi_{\hat{\eta}}^\Seen,
\end{equation}
where $\PP_nf=n^{-1}{\sum_{i=1}^n}f(X_i)$ is the expectation 
$n^{-1}{\sum_{i=1}^n}
f(X_i)$ of a function $f$ under the empirical measure $\PP_n$.
It is reasonable to assume that $(\PP_n-P_{\eta})\dot\chi_{\hat
{\eta}}^\Seen$ is
asymptotically
equivalent to $(\PP_n-P_{\eta})\dot\chi_{{\eta}}^\Seen$ up to the
order $o_P(n^{-1/2})$, as the
difference $(\PP_n-P_{\eta})\dot\chi_{\hat{\eta}}^\Seen$ is ``centered''
and ought to have ``variance'' of the order $O(1/n)$.
(We put ``centered'' and ``variance'' in quotes because
the randomness in the initial estimator $\hat{\eta}$ prevents
a simple calculation of mean and variance.)
Thus, under reasonable regularity conditions the
corrected plug-in estimator (\ref{EqPlugInEstimatorLinear}) will
satisfy
%
%
\begin{eqnarray}
\label{EqLinearEstimatorExpansion}
\quad&&T_n-\chi({\eta})
\nonumber
\\
&&\quad=\chi(\hat{\eta})-\chi({
\eta})+P_{\eta}\dot \chi_{\hat{\eta}}^\Seen+ (
\PP_n-P_{\eta})\dot\chi_{\hat{\eta}}^\Seen
\\
&&\quad=O \bigl(\|\hat{\eta}-{\eta}\|^2 \bigr)+ (\PP_n-P_{\eta})
\dot \chi_{{\eta}}^\Seen+o_P\bigl(n^{-1/2}
\bigr).
\nonumber
\end{eqnarray}
If the first term on the right is sufficiently small, specifically
$\|\hat{\eta}-{\eta}\|=o_P(n^{-1/4})$, then $T_n$ satisfies (\ref{EqASLINEAR})
with $\dot\chi_{\eta}^\Seen$ as the influence function.

The improvement of
the estimator (\ref{EqPlugInEstimatorLinear}) over the ordinary
plug-in estimator is that the estimation error $\|\hat{\eta}-{\eta}\|
$ need
have order $O_P(n^{-1/4})$ rather than $O_P(n^{-1/2})$ for the
estimator to have error $O_P(n^{-1/2})$.
For small ``parametric'' models this is not very relevant, but for semi-
or nonparametric models
the gain can be substantial. For instance, if $\hat{\eta}$ involves an
ordinary smoothing estimator of a regression function on a
$d$-dimensional domain, then a typical rate of estimation is
$n^{-{\alpha}/(2{\alpha}+d)}$, for ${\alpha}$ the number of
derivatives of the true
regression function. This is never $O_P(n^{-1/2})$, but $O_P(n^{-1/4})$ for
${\alpha}\ge d/2$.

The function $\dot\chi_{\eta}^\Seen$ in the von Mises representation
(\ref{EqVonMisesExpansionLinear}) is
exactly an ``influence function'' as in the theory of semiparametric
models (see \cite{KoshevnikLevit,Pfanzagl82,vdV88,BKRW}) and can be
related to the ``tangent set''. Informally, a
\emph{tangent set}
(at $P_{\eta}$) of a model $(P_{\eta}\dvtx {\eta}\in\H)$ is the
set of all
\emph{score functions} at $t=0$,
%
%
\begin{equation}
\label{EqScoreFunction} \dot g_{\eta}:=\frac{\partial}{\partial t}_{|t=0}\log
p_{{\eta}_t} = \biggl(\frac{\partial}{\partial t}_{|t=0}p_{{\eta}_t}
\biggr)\big/{p_{\eta}},
\end{equation}
of (smooth) one-dimensional submodels $(P_{{\eta}_t}\dvtx t\ge0)$
with ${\eta}_0={\eta}$. [Here $t\mapsto{\eta}_t$ is a map from a
neighbourhood of
$0 \in\RR$ to $\H$ such that the derivative (\ref{EqScoreFunction}) exists.]
An \emph{influence function} [of the real parameter $\chi({\eta})$ at
$P_{\eta}$] is defined as a measurable map $x\mapsto\dot\chi_{\eta
}^\Seen(x)$ such that,
for all paths $t\mapsto{\eta}_t$ considered,
%
%
\begin{equation}
\label{EqInfluenceFunctionLinear} \frac{d}{dt}_{|t=0}\chi({\eta}_t)=P_{\eta}
\dot\chi_{\eta}^\Seen \dot g_{\eta}.
\end{equation}
Combining
(\ref{EqTaylorChiLinear})--(\ref{EqVonMisesExpansionLinear}) (with
${\eta}_t$ in the role of ${\eta}$ and ${\eta}$ in the role of $\hat
{\eta}$), we see
that $\chi({\eta}_t)$ is to the first order given by
$\chi({\eta})+P_{{\eta}_t}\dot\chi_{{\eta}}^\Seen$.
Since, according to (\ref{EqScoreFunction}), $\dot g_{\eta} \,
dP_{\eta
}$ is the derivative at $t=0$ of
$dP_{{\eta}_t}$, we next conclude that the function $\dot\chi_{\eta
}^\Seen$ in
the von Mises expansion (\ref{EqVonMisesExpansionLinear}) is an
influence function also in the sense of (\ref{EqInfluenceFunctionLinear}).

An influence function is not necessarily unique,
as only its inner products with elements $\dot g_{\eta}$ of the tangent
set matter. An influence function that is contained
in the closed linear span of the tangent set is called
the \emph{efficient influence function}.
It minimizes the variance $\var_{\eta}\PP_n\dot\chi_{\eta}^\Seen
$ over all
influence functions and is the influence function of asymptotically
efficient estimators.

The theory developed by Robins et al.
in \cite{RobinsetalFreedman} extends the preceding from linear
to higher order approximations.
The motivation is that the parameter ${\eta}$ may be so high
dimensional that no estimator
$\hat{\eta}$ attains the rate $O_P(n^{-1/4})$. The preceding suggests that
then the corrected plug-in estimator will be suboptimal,
as in the expansion (\ref{EqLinearEstimatorExpansion})
the ``bias'' $\chi(\hat{\eta})-\chi({\eta})+P_{\eta}\dot\chi
_{\hat{\eta}}^\Seen$
dominates the ``variance'' $(\PP_n-P_{\eta})\dot\chi_{\hat{\eta
}}^\Seen$.
For this situation Robins et al. \cite{RobinsetalFreedman} introduced
higher order expansions and
influence functions, as follows.

A \emph{tangent set of order $m$} (at $P_{\eta}$) are all
derivatives of the type, for given one-dimensional submodels
$(P_{{\eta}_t}\dvtx t\ge0)$,
%
%
\begin{eqnarray}
\label{EqMScore}
&&\dot g_{\eta}(x_1,\ldots, x_m)\nonumber
\\
&&\quad=
\Biggl(\frac{\partial^j}{\partial
t^j}_{|t=0}{\prod_{i=1}^m}
p_{{\eta}_t}(x_i)\Biggr)\bigg/ \Biggl({\prod
_{i=1}^m}p_{{\eta}}(x_i)\Biggr),
\\
\eqntext{\quad j=1,2,\ldots, m.}
\end{eqnarray}
The functions on the right-hand side are \emph{higher order score
functions} (\cite{smallmcleish,Lindsay}). These are defined
relative to the joint
density $(x_1,\ldots,x_m)\mapsto{\prod_{i=1}^m}p_{\eta}(x_i)$ of $m$
observations, not
as higher order derivatives of a single density, because
higher order derivatives
of the log likelihood of $n$ observations do not reduce to sums over
single observations,
as do first order derivatives. The relationship between expansions on a
single observation and
the joint likelihood can be seen from
\begin{eqnarray*}
&&{\prod_{i=1}^n}\frac{p_{{\eta}_t}}{p_{\eta}}(x_i)
\\
&&\quad={\prod_{i=1}^n} \biggl(1+t \dot
g_{\eta}(x_i)+\thalf t^2\ddot
g_{\eta}(x_i) +\cdots \biggr)
\\
&&\quad= 1+t {\sum_{i=1}^n}\dot
g_{\eta}(x_i)
\\
&&\qquad{}+t^2 \Biggl(\thalf{\sum
_{i=1}^n}\ddot g_{\eta}(x_i) +
\mathop{\sum\sum}\limits
_{1\le i<j\le n} \dot g_{\eta}(x_i)\dot
g_{\eta}(x_j) \Biggr)
\\
&&\qquad{}+\cdots.
\end{eqnarray*}
Inspection of this expansion shows that the coefficient of
$t^j$ is a $U$-statistic of degree $j$ [cf. equation
(\ref{EqUstatistic}) below]. The kernels of these
$U$-statistics up to order $m$ can also be obtained as
higher order derivatives of products of $m$ densities, as in
(\ref{EqMScore}). Furthermore, they are \emph{degenerate} in the
sense that the integral of a kernel with respect to a single coordinate relative
to the true density $p_{\eta}$ is
zero, generalizing the property that a score function has mean zero;
equivalently, this property can be described as orthogonality of higher
order score functions relative to lower order score functions.

Correspondingly, an \emph{influence function of order $m$} [of the map
${\eta}\mapsto\chi({\eta})$ at $P_{\eta}$] is a measurable map
$(x_1,\ldots, x_m)\mapsto
\dot\chi_{\eta}(x_1,\ldots, x_m)$ such that, for every given
one-dimensional submodel
$(P_{{\eta}_t}\dvtx t\ge0)$,
%
%
\begin{equation}
\label{EqHIgherOrderIF} \qquad\frac{\partial^j}{\partial t^j}_{|t=0}\chi(p_{{\eta}_t})=
P_{{\eta}}^m\dot\chi_{{\eta}}\dot g_{\eta},\quad
j=1,2,\ldots, m.
\end{equation}
This influence function is determined only up to its inner products
with the tangent set and hence is not unique. A minimal version could
be defined as one such that the variance of the $U$-statistic with
kernel $\dot\chi_{\eta}$ is minimal.

For computation in examples the defining equations (\ref{EqHIgherOrderIF})
of a higher order influence function can be tedious. It is usually
easier to apply the rule that a higher order derivative is
the derivative of the previous order derivative
(as shown for second order influence functions in \cite{Pfanzagl85},
4.3.11). One computes the first order influence function
$x_1\mapsto\dot\chi_{\eta}^\Seen(x_1)$ of the functional ${\eta
}\mapsto\chi({\eta})$ as
usual. Next one recursively for $j=2,3,\ldots, m$
determines
influence functions, written $x_j\mapsto\dot\chi_{\eta}^{\Sj
j}(x_1,\ldots, x_j)$
as influence functions of the functionals ${\eta}\mapsto
\dot\chi_{\eta}^{\Sj{j-1}}(x_1,\ldots, x_{j-1})$,
for fixed $(x_1,\ldots, x_{j-1})$. The function $\dot\chi_{\eta}^j$
can be made
degenerate (in the sense defined previously) by subtracting its projection
on the linear span of all functions of one argument less. Then
\[
\dot\chi_{\eta}(x_1,\ldots,x_m)=\sum
_{j=1}^m \frac{1} {j!} {\dot
\chi_{\eta}^{\Sj j}(x_1,\ldots, x_j)}
\]
is an $m$th order influence function. As we consider only a single value
of $m$ at a time, we do not let $m$ show up in the notation on the left.
As a consequence, the formulas in the following will appear as in the
linear case.

Given an influence function of order $m$, we may now generalize
the definition of the improved plug-in estimator
(\ref{EqPlugInEstimatorLinear}) to
%
%
\begin{equation}
\label{EqPlugInEstimatorM} T_n=\chi(\hat{\eta})+\UU_n \dot
\chi_{\hat{\eta}},
\end{equation}
for $\UU_nf$ denoting a $U$-statistic of order $m$ with kernel $f$:
%
%
\begin{eqnarray}
\label{EqUstatistic} \UU_nf&=&\frac{1}{n(n-1)\cdots(n-m+1)}
\nonumber
\\[-8pt]
\\[-8pt]
&&{}\cdot\sum
_{1\le
i_1\neq i_2\neq\cdots\neq i_m\le n} f(X_{i_1},\ldots, X_{i_m}).\nonumber
\end{eqnarray}
The term $\UU_n \dot\chi_{\hat{\eta}}$ should correct the plug-in
estimator $\chi(\hat{\eta})$ up to order $m$ and, hence,
an argument similar to (\ref{EqLinearEstimatorExpansion})
should give the expansion
%
%
\begin{eqnarray}
\label{EqMEstimatorExpansion} \quad T_n-\chi({\eta}) &=&O \bigl(\|\hat{\eta}-{\eta}
\|^{m+1} \bigr)
\nonumber
\\[-8pt]
\\[-8pt]
&&{}+ \bigl(\UU_n-P_{\eta
}^m
\bigr)\dot\chi_{{\eta}}+o_P\bigl(n^{-1/2}\bigr).\nonumber
\end{eqnarray}
The bias of the plug-in estimator $\chi(\hat{\eta})$ would be
corrected to the
order $O (\|\hat{\eta}-{\eta}\|^{m+1} )$, and good
estimators for
$\chi({\eta})$ exist even in situations where ${\eta}$ is estimable only
with low precision. The only cost would be a slightly larger variance
in the $U$-statistic relative to the empirical measure.

Unfortunately, there is no such free lunch: one cannot seriously
correct bias without seriously increasing the variance. Although
(\ref{EqMEstimatorExpansion}) and the preceding heuristics are
correct, they do not apply, as higher order influence functions
typically do not exist. Besides by
a lack of invertibility of the map ${\eta}\to p_{\eta}$, this is caused
by failure of a higher order von Mises type representation.
Whereas a continuous, linear map $B\dvtx L_2(P_{\eta})\to\RR$, such as
arises from the first derivative $\chi_{\eta}'$ in
(\ref{EqTaylorChiLinear}), is always
representable as an inner product $B(g)=P_{\eta}\dot\chi_{\eta
}^\Seen g$ for
some function $\dot\chi_{\eta}^\Seen$, a continuous,
multilinear map $B\dvtx L_2(P_{\eta})^j\to\RR$
is not necessarily representable as a repeated integral of the type
\begin{eqnarray*}
&&B(g_1,\ldots, g_j)\\
&&\quad=\int\cdots\int g_1(x_1)
\cdots g_j(x_j)
\\
&&\hphantom{\quad=\int\cdots\int}{}\cdot\dot\chi _{\eta}(x_1,
\ldots,x_j) \,dP_{\eta}(x_1)\cdots
\,dP_{\eta}(x_j).
\end{eqnarray*}
The definition (\ref{EqPlugInEstimatorM}) uses such a
``von Mises representation'' in order to estimate the higher derivatives
using the data, by a $U$-statistic.

We must therefore set a more modest aim: correcting the bias
in certain directions only. A key observation is that a multilinear map
on a
\emph{finite-dimensional} subspace $L\times\cdots\times L\subset
L_2(P_{\eta})^m$
is always representable by a kernel.
If the invertibility ${\eta}\mapsto p_{\eta}$ can be resolved, we can
therefore always ``represent'' and estimate the $m$th order derivative
at differences ${\eta}-\hat{\eta}$ \emph{within} a given
finite-dimensional linear
space. The bias in nonrepresented directions then remains,
and the challenge is to determine the directions that balance three
terms:
\begin{itemize}
\item the bias in the nonrepresented directions,
\emph{representation bias},
\item the estimation error $O_P (\|\hat{\eta}-{\eta}\|
^{m+1} )$, the
\emph{estimation bias},
\item the variance of the resulting $U$-statistic.
\end{itemize}
Regarding the third component, we note that, although the variance
of a $U$-statistic with a fixed kernel is dominated by its linear term
and is of order $O(1/n)$, the need to represent the functionals in
more and more directions given larger sample size $n$ results in
kernels that become more and more complex with $n$. The resulting
variance of $\UU_n\dot\chi_{\hat{\eta}}$ is therefore typically larger
than $O(1/n)$. A new balance should be found with the squared biases,
which will also
be larger than parametric.

The preceding heuristic scheme is general, but its implementation
requires finding the appropriate influence functions that create
the correct bias-variance trade-off. Robins et al.
\cite{RobinsetalFreedman}
achieved this for estimating a functional in a class of
high-dimensional semiparametric
models that includes some popular models for missing data or causal
inference. The high dimensions arise by the inclusion of a
multivariate ``control covariate''. The models have a technical
characterization, through a certain form of the first order influence function.
They are \emph{structured semiparametric models} in that their natural
parameterization is in terms of three or more parameters, which vary
independently.
Thus, the full parameter takes the form ${\eta}=(a,b,c,f)$,
that is partitioned in three subparameters $a$, $b$, $c$ and $f$. The
parameter $f$
is the marginal density of an observable covariate $Z$.
The technical characterization is that the first order influence
function of the parameter of interest
${\eta}\mapsto\chi({\eta})$ can be written in the form
%
%
\begin{eqnarray}
\label{EqJamiesModel} \dot\chi_{\eta}^1(x)&=&a(z)b(z)
S_1(x)+a(z) S_2(x)
\nonumber
\\[-8pt]
\\[-8pt]
&&{}+ b(z) S_3(x)+S_4(x)-
\chi({\eta}),\nonumber
\end{eqnarray}
for known functions $S_i(x)$ of the data [i.e., $S=(S_1,S_2,S_3,S_4)$
is a given statistic]. The covariate $Z$ is assumed to range over a compact
$d$-dimensional domain and the parameters $a,b,f$ are unknown
functions on this domain, restricted only nonparametrically by
smoothness assumptions. The parameter $c$ is an additional parameter
to complete the identification of the distribution of $X$, but it does not
appear in (\ref{EqJamiesModel}).

As the higher order corrections are based on von Mises representations of
higher order influence functions, which are derivatives of the first order
influence function, it is not unnatural to base a theory on the form of the
first order influence function. However, by itself (\ref{EqJamiesModel}) appears
not insightful. The following examples illustrate the class of models.

%
\begin{example}[(Missing data)]
In a version of the \emph{missing data problem} we observe the triple
$X=(YA,A,Z)$, where $Y$ and $A$ are random variables that take values
in the
two-point set $\{0,1\}$ that are conditionally independent
given the variable $Z$. We can think of $Y$ as a response,
which is observed only if the indicator $A$ takes
the value $1$. To ensure independence of the response and
missingness, the covariate $Z$ would be chosen such that
it contains all information on the dependence between $Y$ and $A$
(``missing at
random''). Alternatively, we can think of
$Y$ as a counterfactual outcome if a treatment were
given ($A=1$) and estimate (half) the treatment effect under
the assumption of ``no unmeasured confounders''.
Both applications may require that $Z$ is
high dimensional (e.g., of dimension 10), where there is
typically insufficient a priori information to
model the form of the dependence of $A$ and $Y$ on $Z$.
The three parameters are the marginal density
$f$ of $Z$ and the (inverse) probabilities $b(z)=\Pr(Y=1\given Z=z)$
and $a(z)^{-1}=\Pr(A=1\given Z=z)$. The functional of interest
is the mean response $\E Y$, that is,
\[
\chi({\eta})=\int bf \,d\n.
\]
The representation (\ref{EqJamiesModel}) can be shown to be
valid with $S_1=-A$, $S_2=AY$, $S_3=1$ and $S_4=0$ (see,
e.g., \cite{RobinsetalMetrika}).
The parameters $a$ and $b$ are (transformed) regression functions and
are nonparametrically estimable
at the rates $n^{-{\alpha}/(2{\alpha}+d)}$ and $n^{-\b/(2\b+d)}$
if they are a priori known to be ${\alpha}$- and $\b$-smooth,
where $d$ is the dimension of $Z$. The parameter $f$ is a density
and can be estimated from the covariates. Closer inspection
[see (\ref{EqMARBias}) below] shows that a more crucial parameter
is the quotient $f/a$, which is proportional to the conditional density
of $Z$ given $A=1$
and can be estimated directly from the observed covariates and
treatment indicators,
at a rate $n^{-\g/(2\g+d)}$ if this function is known to be $\g$-smooth.
The purpose of constructing higher order influence functions is to ensure
that standard nonparametric regression or density estimators can
replace the
unknown parameters in theoretical expressions with optimal estimators as
a result.
\end{example}

%
\begin{example}[(Covariance model)]
Let a typical observation be a triple $X=(Y,A,Z)$, where $Y$ and $A$
are binary
variables with values in $\{0,1\}$. We are interested in estimating the
expected conditional product moment $\E [\E(Y\given Z) \E
(A\given Z) ]$. In terms of
the parameters $a(Z)=\E(A\given Z)$ and $b(Z)=\E(Y\given Z)$, and
${\eta}=(a,b,f,c)$, for $f$ the marginal density of $Z$ and $c$ an
additional parameter,
this target can be written as
\[
\chi({\eta})=\int abf \,d\nu.
\]
Representation (\ref{EqJamiesModel}) can be seen to hold
with $S_1=-1$, $S_2=A$, $S_3=Y$ and $S_4=0$. The parameters $a$ and $b$
are regression functions of $Y$ and $A$ on $Z$ and hence can be
estimated at
the rates $n^{-{\alpha}/(2{\alpha}+d)}$ and $n^{-\b/(2\b+d)}$
if they are a priori known to be ${\alpha}$- and $\b$-smooth. The
marginal density
$f$ can be similarly estimated nonparametrically from the observed covariates.

The triple $(a,b,f)$ does not fully parameterize the joint distribution
of an
observation, but the remaining part $c$ of the parameter does not seem
to play a role when estimating
$\chi({\eta})$. A full parameterization is obtained by adding the
\emph{treatment effect function}
$c(Z)=\E(Y\given A=1,Z)-\E(Y\given A=0,Z)$. The conditional distribution
of $Y$ given $A$ can then be expressed in $(a,b,c,f)$ through
$\Pr(Y=1\given A,Z)=c(Z)(A-a(Z))+b(Z)$.

Estimating $\chi({\eta})$ is relevant to the biostatistical setup
through a detour, which
relates $\chi({\eta})$ to the treatment effect function $c$.
First, in terms of statistical difficulty, the functional $\chi({\eta
})$ is equivalent to the functional
$\E\cov(Y,A\given Z)=\E(YA)-\chi({\eta})$, as $\E(YA)$ can be estimated
at the rate $n^{-1/2}$ by a simple sample average. Second, the problem
of estimating $\E\cov(Y,A\given Z)$ is a template for estimating
$\psi(t):=\E\cov(Y-t A,A\given Z)$, for every given $t$, which can
next be inverted
to give an estimate for the value $\t$ that satisfies $\psi(\t)=0$.
The latter
value can be shown to be equal to the \emph{variance weighted average
treatment effect}
\[
\t=\frac{\E\var(A\given Z)c(Z)}{\E\var(A\given Z)}.
\]
(See \cite{RobinsetalMinimax}, Section~4 for details.)
Under the assumption of nonconfounding this parameter
is nonzero if and only if the treatment $A$ has a nonzero causal
effect, and
it may be the ultimate purpose to ascertain this.
\end{example}

%
\begin{example}[(Average treatment effect)]
Suppose a clinical trial with two possible treatments, indicated by
$A\in\{0,1\}$,
has two binary outcome variables $Y_1$ and $Y_2$, and let
$a_j(Z)=\E(Y_j\given A=1,Z)-\E(Y_j\given A=0,Z)$ be
the treatment effects at level $Z$ of an observed covariate, for $j=1,2$.
We observe a random sample of the variables $(Y_1,Y_2,A,Z)$ and are
interested in estimating the
\emph{average treatment effect}
\[
\chi({\eta})=\int a_1a_2 f \,d\nu.
\]
Here ${\eta}$ parameterizes the distribution of $(Y_1,Y_2,A,Z)$, and
$f$ is the density of the covariate $Z$, relative to some measure $\nu$,
for instance, the Lebesgue measure on a compact subset of $\RR^d$.
The parameter ${\eta}$ includes the triplet $(a_1,a_2,f)$ and possibly
other unknown
aspects of the distribution of an observation. In a clinical trial the
probability
$\pi(Z)=\Pr(A=1\given Z)$ that an individual
with covariate $Z$ is treated will be a known function of the covariate.

As the tangent space is a true subspace of the full tangent space,
there are multiple influence functions for $\chi$. It can be shown that
any influence function of $\chi$ can be represented in the form
(\ref{EqJamiesModel}) with, for some measurable function $C$,
\begin{eqnarray*}
S_1&=&1-\frac{2A(A-\pi(Z))}{\pi(Z)(1-\pi)(Z)},
\\
 S_2&=&Y_2
\frac{A-\pi(Z)}{\pi(Z)(1-\pi)(Z)},
\\
S_3&=&Y_1\frac{A-\pi(Z)}{\pi(Z)(1-\pi)(Z)},
\\
 S_4&=&C(Z)
\frac{A-\pi(Z)}{\pi(Z)(1-\pi)(Z)}.
\end{eqnarray*}
Perhaps the special case that $Y_1=Y_2$ is of most interest. The
parameter $(a_1,a_2,f)$
then reduces to a pair $(a,f)$, and $S_2=S_3$, but the general setup
remains the same.
\end{example}

In models with first order influence function of the form
(\ref{EqJamiesModel}) the error of the first order von Mises
representation
(\ref{EqTaylorChiLinear})--(\ref{EqVonMisesExpansionLinear}) can be computed
to be, for a given initial estimator $\hat{\eta}=(\hat a,\hat b,\hat f)$,
%
%
\begin{eqnarray}
\label{EqMARBias}
&&\chi(\hat{\eta})-\chi({\eta}) + P_{\eta}\dot
\chi_{\hat{\eta}}^1
\nonumber
\\[-8pt]
\\[-8pt]
&&\quad=\int(\hat a-a) (\hat b-b) \tilde
s_{{\eta},1} f \,d\n,\nonumber
\end{eqnarray}
for $\tilde s_{{\eta},i}(z)=\E_{\eta}(S_i\given Z=z)$. [From the
fact that
$a$, $b$ and $f$ are only nonparametrically restricted and that
(\ref{EqJamiesModel}) gives the influence function, it can be shown that
necessarily $\tilde s_{{\eta},1} b+\tilde s_{{\eta},2}=0=\tilde
s_{{\eta},1}
a+s_{{\eta},3}$, after which identity (\ref{EqMARBias}) follows by algebra.]
This is quadratic in the errors $\hat a-a$ and $\hat b-b$ of the
initial estimators, but is special in that the squares of the
estimation errors
$|\hat a-a|$ and $|\hat b-b|$ of
the two initial estimators $\hat a$ and $\hat b$ do no arise, but
only their product. This property, termed
``double robustness'' in
\cite{RobinsRotnitzky,RobinsRotnitzkyRegressionMissing}, makes that in
first order
inference it suffices
that one of the two parameters is estimated well. If
initial estimators of $a$ and $b$ attain estimation rates $n^{-{\alpha
}/(2{\alpha}+d)}$ and
$n^{-\b/(2\b+d)}$, respectively, then the order of the remainder term
in the
expansion is the product of these rates.
This shows that the linear estimator (\ref{EqPlugInEstimatorLinear})
attains a
rate $O_P(n^{-1/2})$ if
%
%
\begin{equation}
\label{EqEnoughSmoothnessForRootnByLinear} \frac{\alpha}{2{\alpha}+d}+\frac\b{2\b+d}\ge\frac{1}2.
\end{equation}
If this condition fails, then the ``bias'' (\ref{EqMARBias}) is greater
than $O_P(n^{-1/2})$. The linear
estimator (\ref{EqPlugInEstimatorLinear}) then does not balance bias
and variance and is suboptimal.

For moderate to large dimensions $d$, inequality
(\ref{EqEnoughSmoothnessForRootnByLinear}) is a restrictive
requirement, whose validity is questionable for many applications.
Higher order influence functions allow to construct better estimators
than the linear estimator (\ref{EqPlugInEstimatorLinear}).
As shown in
\cite{RobinsetalFreedman,RobinsetalMetrika,RobinsetalIntegral,RobinsetalMinimax,RobinsetalSPL},
there are two cases:
\begin{itemize}
\item$({\alpha}+\b)/2\ge d/4$. In this case estimation at rate
$n^{-1/2}$ is
possible by using a higher order estimator (\ref{EqPlugInEstimatorM}) of
sufficiently large order $m$. If the inequality is strict, then this
estimator is also semiparametrically regular and efficient, even
though (\ref{EqEnoughSmoothnessForRootnByLinear}) need not be satisfied.
\item$({\alpha}+\b)/2< d/4$. In this case the minimax rate of estimation
is slower than $n^{-1/2}$. If the function $\tilde s_{{\eta},1} f$ has a
regularity $\g$ bigger than a certain cutoff [that depends on
$({\alpha},\b)$], then the minimax rate is $n^{-(2{\alpha}+2\b
)/(2{\alpha}+2\b+1)}$ and
is attainable by a higher order estimator (\ref{EqPlugInEstimatorM}) with
a carefully constructed approximate influence function $\dot\chi
_{\eta}$.
\end{itemize}
In both cases it is necessary to estimate the marginal density
$f$, or rather the function $\tilde s_{{\eta},1}f$, notwithstanding
the fact that it does not enter the
first order influence function (\ref{EqJamiesModel}).
Robins et al. \cite{RobinsetalFreedman} construct minimax estimators under
the assumption that this function has a minimal smoothness.
A completely general solution is apparently still more complicated.

The details of the constructions are beyond the scope of the present paper.
The approximations are based on expanding the
parameters $a$ and $b$ on bases that express their regularity (e.g.,
suitable wavelets)
and representing the higher order derivatives of the functional $\chi$
on the subspaces obtained
by truncating these bases. The truncation point is chosen relative to the
functional to be estimated (and not necessarily the usual one used to
estimate the functions
themselves). For orders three and up, it is in addition necessary to remove
pairs of basis functions [resulting from the pair $(a,b)$] whose combined
index is ``large'', in order to cut variance without increasing bias.
For an introduction to constructing truncated second order influence
functions we refer to
\cite{RobinsetalMetrika}.

\section{Concluding Remarks}

One may look at the work of Robins et al. \cite{RobinsetalFreedman}
and its sequel
from two perspectives. The mathematical
statistical point of view is the simplest: higher order estimating functions
are a means to construct estimators that are theoretically minimax in complex
semiparametric models, where the interest is not simply in a mean of
the observations,
but in a parameter defined through the structure of the model. As always
in high-dimensional models, minimaxity is about the bias-variance trade-off.
Inspection of higher order tangent spaces reveals in what form the bias
arises, and
the connected von Mises calculus allows to correct for it.
So far no completely general method exists for trading this against
variance (other than
the abstract idea to use ``finite-dimensional approximations''), and, in
fact, beyond the
application to models characterized by (\ref{EqJamiesModel}), nothing
much is known.

The second perspective is practically oriented. The models dealt with
in this paper are relevant in
studies in epidemiology, econometrics and the social sciences. The
parameter of interest is
defined through the substantial application, for instance, measuring a
response to treatment or the
consequence of an intervention. High dimensions arise to identify this
parameter of interest
from data. Observational studies, where covariates must be included in the
statistical analysis to control for possible confounding, are a typical
case. One has a choice to
adopt a relatively simple statistical model for this complex reality,
maybe even a classical
parametric model or a one-dimensional propensity score, or to let the
data ``speak for itself'', as
much as possible. Without any model restriction one runs into the
``curse of dimensionality'' and no
conclusions are possible. Semiparametric models as developed in the
1980s and 1990s are between
these extremes, but from the present perspective relatively close to
finite-dimensional models. In
fact, they focus on functionals in situations where a bias-variance
trade-off is unnecessary, as the
bias is negligible. The main purpose of methods based on
high-dimensional influence functions is to
fill the huge gap between ``classical semiparametric models'' and the
model in which nothing is
assumed. In a situation with fewer or less stringent a priori
assumptions on the model, statistical
bias starts playing a role and must be traded versus variance.
Estimators with bigger standard
errors result, but bias due to model misspecification decreases. The
choice between model
bias with smaller variance and larger estimation variance is not easy
to make with current
statistical methodology. However, larger and larger data bases
certainly make the
methodology of higher influence functions feasible.

Thus, these methods are potentially useful to answer a wide range of questions.
We close with some remarks about further research that needs to be done
to make the methods fully operational.

The improved estimators based on higher order influence functions
combine good preliminary estimators for deviations of the parameter
of interest $\chi({\eta})$ in some directions with a priori assumptions
that the deviations in other ``nonestimable'' directions are small.
The latter a priori assumptions are always questionable. It is an open
problem to develop estimation procedures that can ``adapt'' to ``scales of
a priori conditions'', for instance, by implicitly estimating unknown
smoothness levels
from the data.

For practical application, estimation without error indications are
insufficient. Although there is some preliminary work on confidence
intervals related to the higher order estimators, these procedures
remain to be explored.

The models (\ref{EqJamiesModel}) considered in \cite
{RobinsetalFreedman} are structured
semiparametric models [with a partitioned parameter $(a,b,c,f)$ and the
functional
of interest defined naturally in terms of the partition], but
typically nonparametric in the sense that any law on the sample space
is realized by some choice of the parameters $(a,b,c,f)$.
Genuinely semiparametric problems, such as partial linear
regression, pose a further challenge. For such models the first order influence
function is nonunique and, as the estimation error is bigger than the
first order variance, the efficient first order influence function may
not play a special role, thus increasing the degrees of freedom in
constructing suitable higher order influence functions.

\section*{Acknowledgment}

Supported in part from the European
Research Council under ERC Grant Agreement 320637.


%
%


\begin{thebibliography}{17}

\bibitem{BegunHallHuangWellner}
%
\begin{barticle}[mr]
\bauthor{\bsnm{Begun},~\bfnm{Janet~M.}\binits{J.~M.}},
\bauthor{\bsnm{Hall},~\bfnm{W.~J.}\binits{W.~J.}},
\bauthor{\bsnm{Huang},~\bfnm{Wei-Min}\binits{W.-M.}} \AND
\bauthor{\bsnm{Wellner},~\bfnm{Jon~A.}\binits{J.~A.}}
(\byear{1983}).
\btitle{Information and asymptotic efficiency in
parametric--nonparametric models}.
\bjournal{Ann. Statist.}
\bvolume{11}
\bpages{432--452}.
\bid{doi={10.1214/aos/1176346151}, issn={0090-5364}, mr={0696057}}
\end{barticle}
%
\bptok{imsref}%
\endbibitem

\bibitem{BKRW}
%
\begin{bbook}[mr]
\bauthor{\bsnm{Bickel},~\bfnm{Peter~J.}\binits{P.~J.}},
\bauthor{\bsnm{Klaassen},~\bfnm{Chris~A.~J.}\binits{C.~A.~J.}},
\bauthor{\bsnm{Ritov},~\bfnm{Ya'acov}\binits{Y.}} \AND
\bauthor{\bsnm{Wellner},~\bfnm{Jon~A.}\binits{J.~A.}}
(\byear{1993}).
\btitle{Efficient and Adaptive Estimation for Semiparametric Models}.
\bpublisher{Johns Hopkins Univ. Press},
\blocation{Baltimore, MD}.
\bid{mr={1245941}}
\end{bbook}
%
\bptok{imsref}%
\endbibitem


\bibitem{vdVStFlour}
%
\begin{bbook}[mr]
\bauthor{\bsnm{Bolthausen},~\bfnm{E.}\binits{E.}},
\bauthor{\bsnm{Perkins},~\bfnm{E.}\binits{E.}} \AND
\bauthor{\bsnm{van~der Vaart},~\bfnm{A.}\binits{A.}}
(\byear{2002}).
\btitle{Lectures on Probability Theory and Statistics}.
\bseries{Lecture Notes in Math.}
\bvolume{1781}.
\bpublisher{Springer},
\blocation{Berlin}.
\bid{mr={1915443}}
\end{bbook}
%
\bptok{imsref}%
\endbibitem

\bibitem{KoshevnikLevit}
%
\begin{barticle}[mr]
\bauthor{\bsnm{Ko{\v{s}}evnik},~\bfnm{Ju.~A.}\binits{Ju.~A.}} \AND
\bauthor{\bsnm{Levit},~\bfnm{B.~Ja.}\binits{B.~Ja.}}
(\byear{1976}).
\btitle{On a nonparametric analogue of the information matrix}.
\bjournal{Teor. Veroyatn. Primen.}
\bvolume{21}
\bpages{759--774}.
\bid{issn={0040-361X}, mr={0428578}}
\end{barticle}
%
\bptok{imsref}%
\endbibitem

\bibitem{RobinsetalSPL}
%
\begin{barticle}[mr]
\bauthor{\bsnm{Li},~\bfnm{Lingling}\binits{L.}},
\bauthor{\bsnm{Tchetgen Tchetgen},~\bfnm{Eric}\binits{E.}},
\bauthor{\bsnm{van~der Vaart},~\bfnm{Aad}\binits{A.}} \AND
\bauthor{\bsnm{Robins},~\bfnm{James~M.}\binits{J.~M.}}
(\byear{2011}).
\btitle{Higher order inference on a treatment effect under low
regularity conditions}.
\bjournal{Statist. Probab. Lett.}
\bvolume{81}
\bpages{821--828}.
\bid{doi={10.1016/j.spl.2011.02.030}, issn={0167-7152}, mr={2793749}}
\end{barticle}
%
\bptok{imsref}%
\endbibitem

\bibitem{Lindsay}
%
\begin{barticle}[mr]
\bauthor{\bsnm{Lindsay},~\bfnm{B.~G.}\binits{B.~G.}}
(\byear{1983}).
\btitle{Efficiency of the conditional score in a mixture setting}.
\bjournal{Ann. Statist.}
\bvolume{11}
\bpages{486--497}.
\bid{issn={0090-5364}, mr={0696061}}
\end{barticle}
%
\bptok{imsref}%
\endbibitem

\bibitem{Pfanzagl82}
%
\begin{bbook}[mr]
\bauthor{\bsnm{Pfanzagl},~\bfnm{Johann}\binits{J.}}
(\byear{1982}).
\btitle{Contributions to a General Asymptotic Statistical Theory}.
\bseries{Lecture Notes in Statistics}
\bvolume{13}.
\bpublisher{Springer},
\blocation{New York}.
\bid{mr={0675954}}
\end{bbook}
%
\bptok{imsref}%
\endbibitem

\bibitem{Pfanzagl85}
%
\begin{bbook}[mr]
\bauthor{\bsnm{Pfanzagl},~\bfnm{J.}\binits{J.}}
(\byear{1985}).
\btitle{Asymptotic Expansions for General Statistical Models}.
\bseries{Lecture Notes in Statistics}
\bvolume{31}.
\bpublisher{Springer},
\blocation{Berlin}.
\bid{doi={10.1007/978-1-4615-6479-9}, mr={0810004}}
\end{bbook}
%
\bptok{imsref}%
\endbibitem

\bibitem{RobinsetalFreedman}
%
\begin{bincollection}[mr]
\bauthor{\bsnm{Robins},~\bfnm{James}\binits{J.}},
\bauthor{\bsnm{Li},~\bfnm{Lingling}\binits{L.}},
\bauthor{\bsnm{Tchetgen},~\bfnm{Eric}\binits{E.}} \AND
\bauthor{\bsnm{van~der Vaart},~\bfnm{Aad}\binits{A.}}
(\byear{2008}).
\btitle{Higher order influence functions and minimax estimation of
nonlinear functionals}.
In \bbooktitle{Probability and Statistics: Essays in Honor of {D}avid
{A}. {F}reedman}.
\bseries{Inst. Math. Stat. Collect.}
\bvolume{2}
\bpages{335--421}.
\bpublisher{IMS},
\blocation{Beachwood, OH}.
\bid{doi={10.1214/193940307000000527}, mr={2459958}}
\end{bincollection}
%
\bptok{imsref}%
\endbibitem

\bibitem{RobinsetalMetrika}
%
\begin{barticle}[mr]
\bauthor{\bsnm{Robins},~\bfnm{James}\binits{J.}},
\bauthor{\bsnm{Li},~\bfnm{Lingling}\binits{L.}},
\bauthor{\bsnm{Tchetgen},~\bfnm{Eric}\binits{E.}} \AND
\bauthor{\bsnm{van~der Vaart},~\bfnm{Aad~W.}\binits{A.~W.}}
(\byear{2009}).
\btitle{Quadratic semiparametric von {M}ises calculus}.
\bjournal{Metrika}
\bvolume{69}
\bpages{227--247}.
\bid{doi={10.1007/s00184-008-0214-3}, issn={0026-1335}, mr={2481922}}
\end{barticle}
%
\bptok{imsref}%
\endbibitem

\bibitem{RobinsRotnitzky}
\begin{barticle}[auto]
\bauthor{\bsnm{Robins},~\bfnm{J.}\binits{J.}} \AND
\bauthor{\bsnm{Rotnitzky},~\bfnm{A.}\binits{A.}}
(\byear{2001}).
\btitle{Comment on the Bickel and Kwon article
``Inference for semiparametric
  models: Some questions and an answer.''}
\bjournal{Statist. Sinica}
\bvolume{11}
\bpages{920--936}.
\end{barticle}
%
\bptok{imsref}%
\endbibitem


\bibitem{RobinsetalMinimax}
%
\begin{barticle}[mr]
\bauthor{\bsnm{Robins},~\bfnm{James}\binits{J.}},
\bauthor{\bsnm{Tchetgen Tchetgen},~\bfnm{Eric}\binits{E.}},
\bauthor{\bsnm{Li},~\bfnm{Lingling}\binits{L.}} \AND
\bauthor{\bsnm{van~der Vaart},~\bfnm{Aad}\binits{A.}}
(\byear{2009}).
\btitle{Semiparametric minimax rates}.
\bjournal{Electron. J. Stat.}
\bvolume{3}
\bpages{1305--1321}.
\bid{doi={10.1214/09-EJS479}, issn={1935-7524}, mr={2566189}}
\end{barticle}
%
\bptok{imsref}%
\endbibitem

\bibitem{RobinsRotnitzkyRegressionMissing}
%
\begin{barticle}[mr]
\bauthor{\bsnm{Robins},~\bfnm{James~M.}\binits{J.~M.}} \AND
\bauthor{\bsnm{Rotnitzky},~\bfnm{Andrea}\binits{A.}}
(\byear{1995}).
\btitle{Semiparametric efficiency in multivariate regression models
with missing data}.
\bjournal{J. Amer. Statist. Assoc.}
\bvolume{90}
\bpages{122--129}.
\bid{issn={0162-1459}, mr={1325119}}
\end{barticle}
%
\bptok{imsref}%
\endbibitem

\bibitem{smallmcleish}
%
\begin{bbook}[mr]
\bauthor{\bsnm{Small},~\bfnm{Christopher~G.}\binits{C.~G.}} \AND
\bauthor{\bsnm{McLeish},~\bfnm{D.~L.}\binits{D.~L.}}
(\byear{1994}).
\btitle{Hilbert Space Methods in Probability and Statistical Inference}.
\bpublisher{Wiley},
\blocation{New York}.
\bid{doi={10.1002/9781118165522}, mr={1269321}}
\end{bbook}
%
\bptok{imsref}%
\endbibitem\

\bibitem{RobinsetalIntegral}
%
\begin{barticle}[mr]
\bauthor{\bsnm{Tchetgen},~\bfnm{Eric}\binits{E.}},
\bauthor{\bsnm{Li},~\bfnm{Lingling}\binits{L.}},
\bauthor{\bsnm{Robins},~\bfnm{James}\binits{J.}} \AND
\bauthor{\bsnm{van~der Vaart},~\bfnm{Aad}\binits{A.}}
(\byear{2008}).
\btitle{Minimax estimation of the integral of a power of a density}.
\bjournal{Statist. Probab. Lett.}
\bvolume{78}
\bpages{3307--3311}.
\bid{doi={10.1016/j.spl.2008.07.001}, issn={0167-7152}, mr={2479495}}
\end{barticle}
%
\bptok{imsref}%
\endbibitem

\bibitem{vanderVaart91}
%
\begin{barticle}[mr]
\bauthor{\bsnm{van~der Vaart},~\bfnm{Aad}\binits{A.}}
(\byear{1991}).
\btitle{On differentiable functionals}.
\bjournal{Ann. Statist.}
\bvolume{19}
\bpages{178--204}.
\bid{doi={10.1214/aos/1176347976}, issn={0090-5364}, mr={1091845}}
\end{barticle}
%
\bptok{imsref}%
\endbibitem

\bibitem{vdV88}
%
\begin{bbook}[mr]
\bauthor{\bsnm{van~der Vaart},~\bfnm{A.~W.}\binits{A.~W.}}
(\byear{1988}).
\btitle{Statistical Estimation in Large Parameter Spaces}.
\bseries{CWI Tract}
\bvolume{44}.
\bpublisher{Stichting Mathematisch Centrum, Centrum voor Wiskunde en
Informatica},
\blocation{Amsterdam}.
\bid{mr={0927725}}
\end{bbook}
%
\bptok{imsref}%
\endbibitem
\end{thebibliography}
\end{document}